\pdfoutput=1

\documentclass[iop]{emulateapj}

\slugcomment{}

\shorttitle{Star formation activities in the molecular cloud associated with S242}
\shortauthors{L.~K. Dewangan}
\begin{document}

\title{The molecular cloud S242: physical environment and star formation activities}
\author{L.~K. Dewangan\altaffilmark{1}, T. Baug\altaffilmark{2}, D.~K. Ojha\altaffilmark{2}, P. Janardhan\altaffilmark{1}, 
R. Devaraj\altaffilmark{3}, and A. Luna\altaffilmark{3}}
\email{lokeshd@prl.res.in}
\altaffiltext{1}{Physical Research Laboratory, Navrangpura, Ahmedabad - 380 009, India.}
\altaffiltext{2}{Department of Astronomy and Astrophysics, Tata Institute of Fundamental Research, Homi Bhabha Road, Mumbai 400 005, India.}
\altaffiltext{3}{Instituto Nacional de Astrof\'{\i}sica, \'{O}ptica y Electr\'{o}nica, Luis Enrique Erro \# 1, Tonantzintla, Puebla, M\'{e}xico C.P. 72840.}
%
\begin{abstract}
We present a multi-wavelength study to probe the star formation (SF) processes on a larger 
scale ($\sim$1$\degr$.05 $\times$ 0$\degr$.56) around the S242 site. The S242 molecular cloud is depicted in a velocity range 
from $-$3.25 to 4.55 km s$^{-1}$ and has spatially elongated appearance. Based on the virial analysis, the cloud is prone to gravitational collapse. 
The cloud harbors an elongated filamentary structure (EFS; length $\sim$25 pc) evident in the {\it Herschel} column density map and the EFS has an observed mass per unit length of $\sim$200 M$_{\odot}$ pc$^{-1}$ exceeding the critical value of $\sim$16 M$_{\odot}$ pc$^{-1}$ (at T = 10 K). 
The EFS contains a chain of {\it Herschel} clumps (M$_{clump}$ $\sim$150 to 1020 M$_{\odot}$), revealing the evidence of fragmentation along its length. 
The most massive clumps are observed at both the EFS ends, while the S242 H\,{\sc ii} region is located at one EFS end. 
Based on the radio continuum maps at 1.28 and 1.4 GHz, the S242 H\,{\sc ii} region is ionized by a B0.5V--B0V type star and has a 
dynamical age of $\sim$0.5 Myr. The photometric 1--5 $\mu$m data analysis of point-like sources traces young 
stellar objects (YSOs) toward the EFS and the clusters of YSOs are exclusively found at both the EFS ends, revealing the SF activities. 
Considering the spatial presence of massive clumps and YSO clusters at both the EFS ends, the observed results are consistent with the prediction of a SF scenario of the end-dominated collapse driven by the higher accelerations of gas. 
 \end{abstract}
\section{Introduction}
\label{sec:intro}
The {\it Herschel} continuum observations demonstrated clearly that the filaments are common features seen in the star-forming regions \citep[e.g.][]{andre10}.
The dust continuum maps have been utilized as a very useful tool to investigate the filaments 
and to infer the underlying structure(s) within the filaments.
These filaments are observed at various scales and often contain the star-forming clumps and cores along their lengths 
\citep[e.g.][and references therein]{schneider12,ragan14,contreras16,li16}. 
However, the physical mechanisms concerning to their formation and their link to the 
star formation processes are not well understood. The role of filaments in the formation of dense massive star-forming clumps and clusters is 
also unknown \citep[e.g.][]{myers09,schneider12,nakamura14,tan14,andre16,kainulainen16}. 
In star-forming regions, the knowledge of physical conditions, kinematics of the molecular gas, and masses per unit length of filamentary features can help us 
to understand the ongoing physical processes.

The star-forming region, LBN 182.30+00.07 or Sh 2-242 (hereafter S242) is situated at a distance of 2.1 kpc \citep{blitz82}. 
The H\,{\sc ii} region linked with the S242 site (hereafter S242 H\,{\sc ii} region) is ionized by a star BD+26 980 of 
spectral type B0V \citep{hunter90}. 
In the S242 H\,{\sc ii} region, \citet{fich90} reported the radial velocity of H$\alpha$ emission to be about $-$0.6 km s$^{-1}$. 
Using the CO line data, \citet{blitz82} estimated the radial velocity of molecular gas to be 0.0$\pm$0.5 km s$^{-1}$ toward S242.  
\citet{kawamura98} also studied the molecular gas content of molecular clouds in Gemini and Auriga including the S242 site
using $^{13}$CO (1-0) emission \citep[see S242 region around $l$ = 182$\degr$.40; $b$ = 0$\degr$.27 in Figures 1, 2, and 9j in][]{kawamura98}. 
They referred the molecular cloud associated with S242 as ``182.4+00.3" cloud (V$_{lsr}$ $\sim$0.7 km s$^{-1}$; line width ($\Delta V$) = 2.1 km s$^{-1}$) 
and estimated the mass of the cloud (M$_{cloud}$) to be $\sim$7000 M$_{\odot}$ \citep[radius $\sim$7 pc; see source ID \#73 in Table 1 in][]{kawamura98}. 
The cloud appears spatially elongated in the 
$^{13}$CO (1-0) map \citep[see Figures 1 and 9j in][]{kawamura98}. However, the identification of filaments and their role in 
star formation processes are yet to be probed within the molecular cloud 182.4+00.3/S242. 
\citet{snell90} investigated a CO outflow toward IRAS 05490+2658 and found that this IRAS source is located $\sim$5$\arcmin$ east of the 
S242 H\,{\sc ii} region. \citet{beuther02} investigated two 1.2 mm peaks toward IRAS 05490+2658 with the IRAM 30-m 
telescope (spatial resolution $\sim$11$\arcsec$). The H$_{2}$ emission at 2.12 $\mu$m is also traced near the IRAS 05490+2658 \citep[see Figure A9 in][]{varricatt10} 
and they also mentioned the presence of two clusters of infrared excess sources near the IRAS 05490+2658. 
Together, these previous studies indicate that the S242 is an active site of star formation and also contains a massive star. 
However, the impact of the massive star in its vicinity is yet to be examined in this star-forming site. 
On a larger scale, the physical conditions around the S242 site are yet to be investigated. 
In this paper, to investigate the ongoing physical mechanisms in and around the S242 site, we carry out a detailed multi-wavelength study of observations from optical, near-infrared (NIR), mid-infrared (MIR), far-infrared (FIR), sub-millimeter (sub-mm), millimeter (mm) to centimeter (cm) wavelengths, including the Giant Metre-wave Radio Telescope (GMRT) radio continuum map at 1.28 GHz and United Kingdom Infra-Red Telescope (UKIRT) Infrared Deep Sky Survey (UKIDSS) NIR data.  

In Section~\ref{sec:obser}, we give the description of the multi-band data-sets utilized in the present work. 
In Section~\ref{sec:data}, the results concerning the physical environment and point-like sources are summarized.  
The possible star formation processes are discussed in Section~\ref{sec:disc} . 
Finally, the main results are summarized and concluded in Section~\ref{sec:conc}.
\section{Data and analysis}
\label{sec:obser}
In this work, we have selected a region of $\sim$1$\degr$.05 $\times$ 0$\degr$.56 ($\sim$38.5 pc $\times$ 20.5 pc at a distance of 2.1 kpc) 
(central coordinates: $l$ = 182$\degr$.217; $b$ = 0$\degr$.239) around the S242 site. 
The target area is selected in such a way that it contains the previously known molecular cloud, ``182.4+00.3". 
In the following, we provide a brief description of the adopted multi-wavelength data-sets. 
\subsection{H$\alpha$ Image}
We retrieved narrow-band H$\alpha$ image at 0.6563 $\mu$m from the 
Isaac Newton Telescope (INT) Photometric H$\alpha$ Survey of the Northern Galactic Plane \citep[IPHAS;][]{drew05} survey database.
The IPHAS imaging survey was carried out using the Wide-Field Camera (WFC) at the 2.5-m INT, located at La Palma. 
The WFC contains four 4k $\times$ 2k CCDs, in an L-shape configuration. The pixel scale is $0\farcs33$ (see \citet{drew05} for more details). 
\subsection{NIR (1--5 $\mu$m) Data}
%
We utilized the NIR photometric magnitudes of point-like sources extracted from the UKIDSS DR10PLUS Galactic Plane Survey \citep[GPS;][]{lawrence07} and the Two Micron All Sky Survey \citep[2MASS;][]{skrutskie06} (hereafter GPS-2MASS). 
The UKIDSS observations (resolution $\sim$$0\farcs8$) were made with the WFCAM mounted on the UKIRT. 
The 2MASS photometric data were utilized to calibrate the UKIDSS fluxes. In this work, we extracted only a reliable NIR photometric catalog. 
More information about the selection procedures of the GPS photometry can be obtained in \citet{dewangan15}.
To obtain reliable 2MASS photometric data, we selected only those sources having photometric magnitude error of 0.1 or less in each band.

Warm-{\it Spitzer} IRAC 3.6 and 4.5 $\mu$m photometric images (resolution $\sim$2$\arcsec$) 
and magnitudes of point sources are downloaded from the Glimpse360\footnote[1]{http://www.astro.wisc.edu/sirtf/glimpse360/} \citep{whitney11} survey.
The photometric magnitudes were extracted from the Glimpse360 highly reliable catalog. 
To obtain further reliable Glimpse360 photometric data, we retrieved only those sources having photometric magnitude error of 0.2 or less in each band.
\subsection{Mid-infrared (12--22 $\mu$m) Data}
MIR images at 12 $\mu$m (spatial resolution $\sim$6$\arcsec$) and 22 $\mu$m (spatial resolution $\sim$1$2\arcsec$) were 
obtained from the publicly available archival WISE\footnote[2]{Wide Field Infrared Survey Explorer, which is a joint project of the
University of California and the JPL, Caltech, funded by the NASA.} \citep{wright10} database.
\subsection{Far-infrared and Sub-millimeter Data}
\label{subsec:her}
We utilized the {\it Herschel} Space Observatory data archives to obtain FIR and sub-mm continuum images. 
The processed level2$_{-}$5 images at 70 $\mu$m, 160 $\mu$m, 250 $\mu$m, 350 $\mu$m, and 500 $\mu$m were 
downloaded using the {\it Herschel} Interactive Processing Environment \citep[HIPE,][]{ott10}. 
The beam sizes of the {\it Herschel} images at 70 $\mu$m, 160 $\mu$m, 250 $\mu$m, 350 $\mu$m, 
and 500 $\mu$m are 5$\farcs$8, 12$\arcsec$, 18$\arcsec$, 25$\arcsec$, and 37$\arcsec$, respectively \citep{poglitsch10,griffin10}. 
In this work, {\it Herschel} temperature and column density maps are produced using the {\it Herschel} continuum 
images, following the methods described in \citet{mallick15}. The {\it Herschel} temperature and column density maps are obtained from a 
pixel-by-pixel spectral energy distribution (SED) fit with a modified blackbody to the cold dust emission at {\it Herschel} 160--500 $\mu$m \citep[also see][]{dewangan15}. The {\it Herschel} 70 $\mu$m data are not included in the analysis, because the 70 $\mu$m emission is 
dominated by the ultraviolet (UV) heated warm dust. Here we provide a brief step-by-step explanation of the procedures. 

The {\it Herschel} 160 $\mu$m image is in unit of Jy pixel$^{-1}$, 
while the images at 250--500 $\mu$m are in units of surface brightness, MJy sr$^{-1}$.
The plate scales of 160, 250, 350, and 500 $\mu$m images are 3$''$.2, 6$''$, 10$''$, 
and 14$''$ pixel$^{-1}$, respectively.  
Prior to the SED fit, the 160--350 $\mu$m images were convolved to the lowest angular 
resolution of 500 $\mu$m image ($\sim$37$\arcsec$) 
and were converted into the same flux unit (i.e. Jy pixel$^{-1}$). 
Furthermore, we regridded these images to the pixel size of 500 $\mu$m image ($\sim$14$\arcsec$). 
These steps were performed using the convolution kernels available in the HIPE software. 
Next, the sky background flux level was estimated to be 0.060, 0.133, 0.198, and $-$0.095 Jy pixel$^{-1}$ for the 500, 350, 250, and 
160 $\mu$m images (size of the selected region $\sim$13$\farcm$4 $\times$ 14$\farcm$8; 
centered at:  $l$ = 183$\degr$.181; $b$ = $-$0$\degr$.354), respectively. 
The negative flux value at 160 $\mu$m is obtained due to the arbitrary scaling of the  {\it Herschel} 160 $\mu$m image.
To avoid diffuse emission linked with the S242 site, the featureless dark field away from the selected target was 
carefully chosen for the background estimation. 

Finally, to generate the temperature and column density maps, a modified blackbody was fitted to the observed fluxes on a pixel-by-pixel basis 
\citep[see equations 8 and 9 in][]{mallick15}. 
The fitting was performed using the four data points for each pixel, maintaining the 
dust temperature (T$_{d}$) and the column density ($N(\mathrm H_2)$) 
as free parameters. 
In the analysis, we used a mean molecular weight per hydrogen molecule ($\mu_{H2}$) of 2.8 
\citep{kauffmann08} and an absorption coefficient ($\kappa_\nu$) of 0.1~$(\nu/1000~{\rm GHz})^{\beta}$ cm$^{2}$ g$^{-1}$, 
including a gas-to-dust ratio ($R_t$) of 100, with a dust spectral index ($\beta$) of 2 \citep[see][]{hildebrand83}. 
We considered flux uncertainties of the order $\sim$15\% in all {\it Herschel} images, based on the previously reported work by \citet{launhardt13}.
We describe the {\it Herschel} temperature and column density maps in Section~\ref{subsec:temp}.
\subsection{Dust continuum 1.1 mm data}
We also obtained Bolocam dust continuum sources at 1.1 mm \citep[v2.1;][]{ginsburg13} from Bolocam Galactic Plane Survey (BGPS). 
The effective full width at half maximum (FWHM) of the 1.1 mm map \citep{aguirre11} is $\sim$33$\arcsec$.
\subsection{Molecular $^{12}$CO line data}
To trace the molecular cloud associated with the S242 site, the 2.6 mm $^{12}$CO data (beam size $\sim$8$\arcmin$)
were obtained from the 1.2-m CfA telescope \citep{dame01}. 
The line data have a velocity resolution of 0.65~km\,s$^{-1}$ and a typical rms value of 0.22 K km s$^{-1}$. 
One can find more details about the $^{12}$CO data in \citet{dame01}. 
\subsection{Radio continuum data}
We utilized the archival radio continuum data at 1.28 GHz and 1.4 GHz.
Radio continuum map at 1.4 GHz (21 cm; beam size  $\sim$45$\arcsec$) was extracted 
from the NRAO VLA Sky Survey \citep[NVSS;][]{condon98} archive, while the radio continuum data of S242 at 1.28 GHz 
were retrieved from the GMRT archive\footnote[3]{https://naps.ncra.tifr.res.in/goa/mt/search/basicSearch}. 
The GMRT observations were obtained on 30 November 2005 (Project Code: 09SKG01). 
The GMRT radio data reduction was carried out using the AIPS software, following similar procedures as described in \citet{mallick13}. 
The 1.28 GHz map has rms noise of 0.15 mJy/beam and beam size of $\sim$20\farcs4 $\times$ 20\farcs4.
\subsection{H\,{\sc i} line data}
We retrieved 21 cm H\,{\sc i} line data from the Canadian Galactic Plane Survey \citep[CGPS;][]{taylor03}.
The velocity resolution of H\,{\sc i} line data is 1.32~km\,s$^{-1}$, sampled every
0.82~km s$^{-1}$. The data have a spatial resolution of 1$\arcmin$ $\times$ 1$\arcmin$ csc$\delta$.
The line data have a brightness-temperature sensitivity of $\Delta$T$_{B}$ = 3.5 sin$\delta$ K.
One can find more details about the CGPS data in \citet{taylor03}.
\section{Results}
\label{sec:data}
\subsection{Large-scale physical environment around S242}
\label{subsec:u1}
In this section, we study the distribution of dense materials, molecular gas, and ionized emission toward S242, enabling us to 
probe the physical environment around the target region. 

\begin{figure*}
\epsscale{1}
\plotone{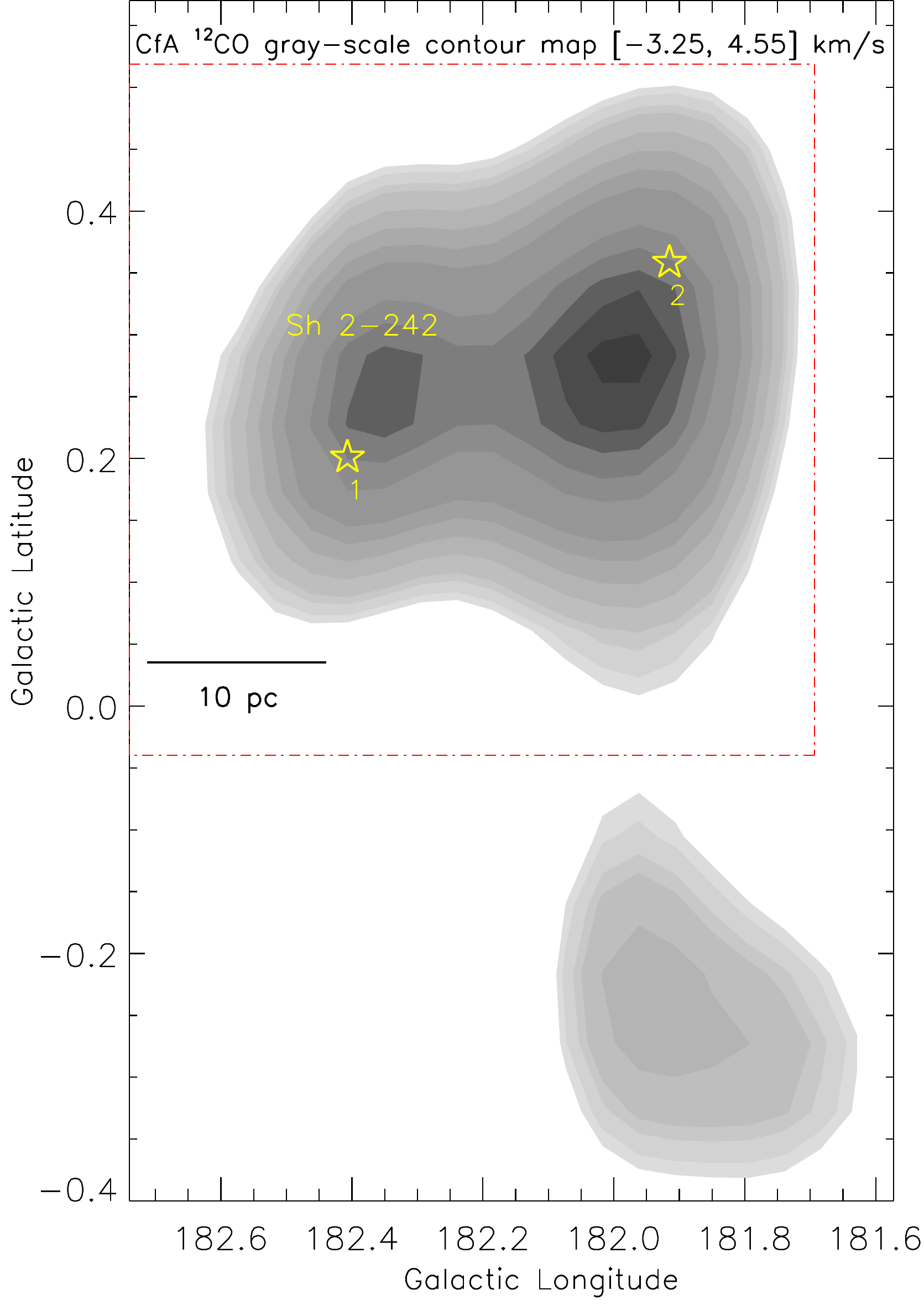}
\caption{\scriptsize The distribution of molecular gas in the direction of our selected field reveals 
an elongated molecular cloud. 
The S242 molecular cloud is depicted in a velocity range from $-$3.25 to 4.55 km s$^{-1}$ and is shown by a gray-scale contour map.
The $^{12}$CO emission contours are shown with levels of 43, 45, 48, 50, 55, 60, 65, 70, 80, 85, 90, 94, and 98\% of 
the peak value (i.e., 10.478 K km s$^{-1}$). The positions of two IRAS sources (IRAS 05488+2657 (\#1) and IRAS 05483+2728 (\#2)) 
are also marked by stars. The scale bar corresponding to 10 pc (at a distance of 2.1 kpc) is shown in the left corner. 
The dotted-dashed red box encompasses the area shown in Figures~\ref{ufg2},~\ref{uf2}a and~\ref{uf2}b. 
The map is smoothed with a Gaussian function with radius of three.}
\label{uf1}
\end{figure*}

\begin{figure*}
\epsscale{1}
\plotone{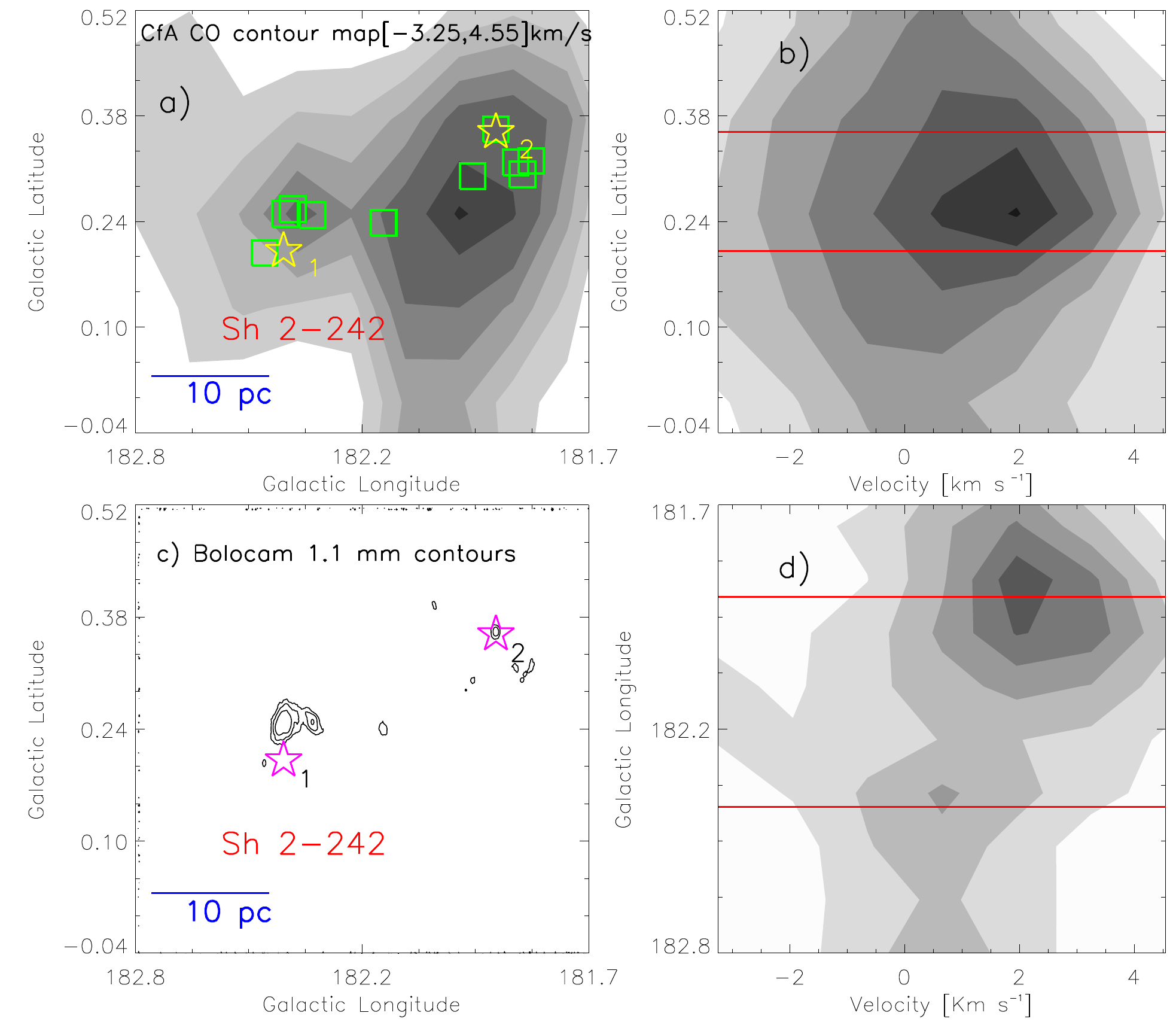}
\caption{\scriptsize a) Integrated intensity map of CfA $^{12}$CO (J = 1-0) around S242. 
The $^{12}$CO emission is shown from $-$3.25 to 4.55 km s$^{-1}$ and is similar to those shown in Figure~\ref{uf1}. 
The positions of dust clumps at 1.1 mm are also marked by square symbols 
and other marked symbols are similar to those shown in Figure~\ref{uf1}.
b) Latitude-velocity distribution of $^{12}$CO. 
The $^{12}$CO emission is integrated over the longitude from 181$\degr$.7 to 182$\degr$.75.
c) Bolocam 1.1 mm dust emission contours are shown with levels of 0.15, 0.3, and 0.6 Jy/beam. 
The other marked symbols are similar to those shown in Figure~\ref{uf1}.
d) Longitude-velocity distribution of $^{12}$CO. The $^{12}$CO emission is integrated over the latitude 
from $-$0.$\degr$04 to 0.$\degr$52. 
In the right panels (i.e., position-velocity maps), two solid red lines show the positions of two IRAS 
sources (IRAS 05488+2657 (\#1) and IRAS 05483+2728 (\#2)).}
\label{ufg2}
\end{figure*}

The study of molecular line data is very essential to trace the physical association of different subregions seen in the large-scale map of a given star-forming region.
Based on the CfA $^{12}$CO line profile, the molecular cloud linked with the S242 site (hereafter S242 molecular cloud) is traced in a velocity range from $-$3.25 to 4.55 km s$^{-1}$ (see Figure~\ref{uf1}). 
Figure~\ref{uf1} shows the spatial distribution of the molecular gas in our selected region around S242, revealing an elongated molecular 
cloud. 
We have also shown the positions of two IRAS sources (i.e. IRAS 05488+2657 and IRAS 05483+2728) in our selected region. 
Based on the distribution of molecular emission, we selected a region of $\sim$1$\degr$.05 $\times$ 0$\degr$.56 containing the S242 site for our present study, 
where a majority of molecular gas have been found (see dotted-dashed box in Figure~\ref{uf1}). 
In Figure~\ref{ufg2}, the integrated $^{12}$CO intensity map (see Figure~\ref{ufg2}a) and the position-velocity maps 
(see Figures~\ref{ufg2}b and~\ref{ufg2}d) are presented.  
In Figure~\ref{ufg2}a, we show the CfA $^{12}$CO intensity map integrated over $-$3.25 to 4.55 km s$^{-1}$. 
The peak positions of ten dust continuum sources at Bolocam 1.1 mm are also marked in Figure~\ref{ufg2}a. 
Bolocam 1.1 mm dust emission contours are presented in Figure~\ref{ufg2}c and trace the cold dust. 
In Figures~\ref{ufg2}b and~\ref{ufg2}d, we show galactic position-velocity diagrams of the $^{12}$CO emission, 
indicating the presence of a single velocity component (at peak velocity $\sim$1.5 km s$^{-1}$) in the direction of our selected target region.

\begin{figure*}
\epsscale{1}
\plotone{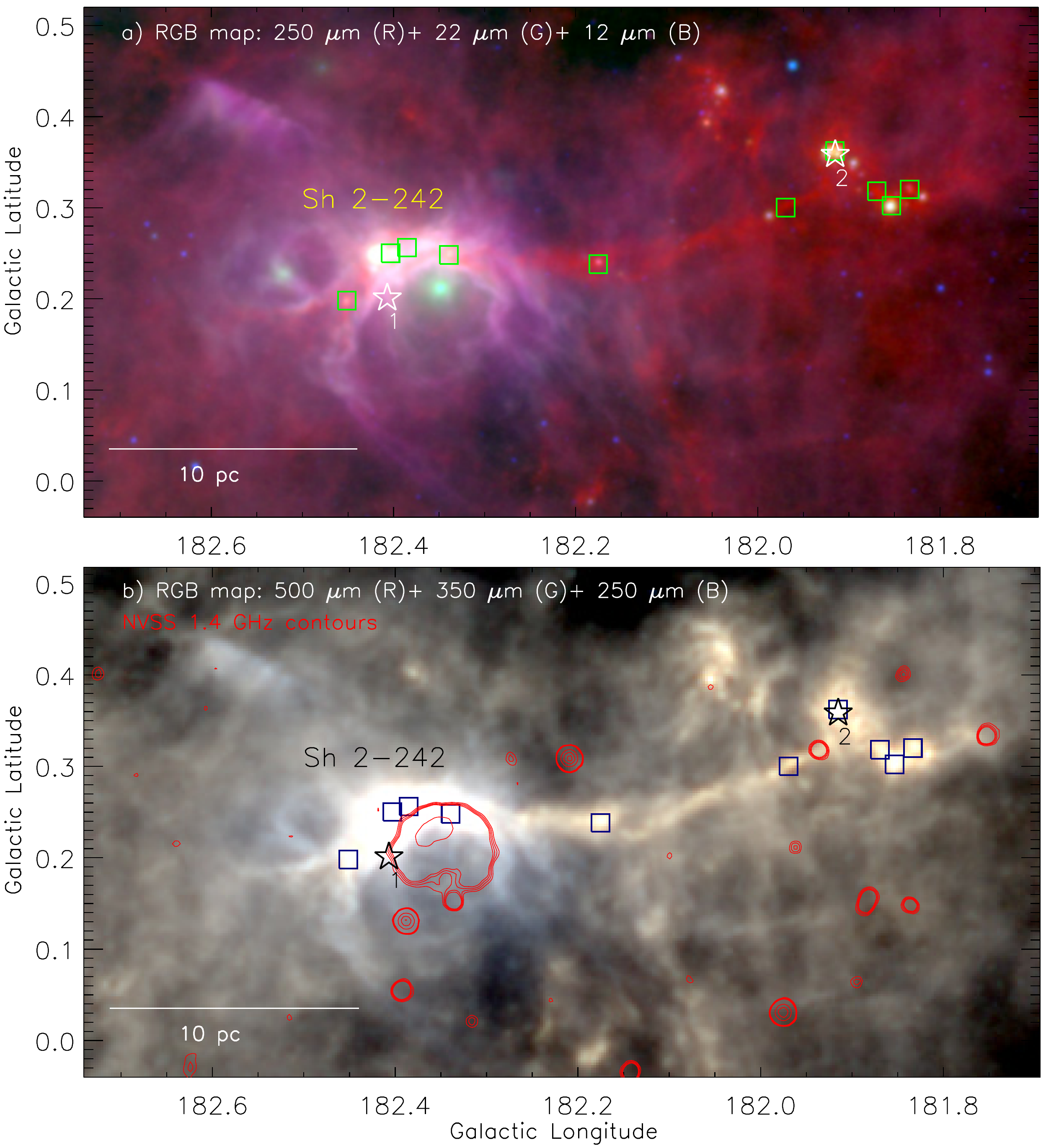}
\caption{\scriptsize A large scale view of the region around S242 
(size of the selected region $\sim$1$\degr$.05 $\times$ 0$\degr$.56 ($\sim$38.5 pc $\times$ 
20.5 pc at a distance of 2.1 kpc); centered at $l$ = 182$\degr$.217; $b$ = 0$\degr$.239).
a) Color composite map is the result of the combination of three bands (in logarithmic scale): 
250 $\mu$m in red ({\it Herschel}), 22 $\mu$m in green ({\it WISE}), and 12 $\mu$m in blue ({\it WISE}). 
b) The distribution of the sub-mm emission toward the region around S242. 
Color composite map is the result of the combination of three {\it Herschel} bands (500 $\mu$m (red), 350 $\mu$m (green), and 250 $\mu$m (blue)) 
and is overlaid with the NVSS 1.4 GHz contours. The 1.4 GHz contours (in red) are superimposed with 
levels of 0.55, 1, 2, 3, 4, 5, 6, 30, 60, and 90\% of the peak value (i.e.  0.377 Jy/beam). 
In both the panels, the positions of dust clumps at 1.1 mm are marked by square symbols and other marked symbols are 
similar to those shown in Figure~\ref{uf1}.}
\label{uf2}
\end{figure*}

Figure~\ref{uf2}a shows a three-color composite map made using the {\it Herschel} 250 $\mu$m in red, {\it WISE} 22 $\mu$m in green, and 
{\it WISE} 12 $\mu$m in blue. The images at wavelengths longer than 150 $\mu$m trace the cold dust emission, 
while the 12--70 $\mu$m emission is sensitive for the warm dust components. 
Figure~\ref{uf2}b shows the {\it Herschel} sub-mm images overlaid with the NVSS 1.4 GHz continuum emission. 
The NVSS map allows to infer the spatial distribution of ionized emission which is only found toward the S242 site.
In Figures~\ref{uf2}a and~\ref{uf2}b, an extended filamentary structure (EFS) (extension $\sim$25 pc; average width $\sim$1.3 pc), containing the S242 site, is revealed and is prominently 
observed in the {\it Herschel} sub-mm images at 250--500 $\mu$m. 
Note that the Bolocam 1.1 mm dust emission map does not trace the entire EFS as seen in 
the {\it Herschel} sub-mm images, however the Bolocam clumps are detected toward the EFS (see Figure~\ref{ufg2}c).
In Figure~\ref{uf1}, with the help of the CfA $^{12}$CO gas distribution, we find a continuous velocity structure in 
the direction of S242 and the EFS is embedded within the S242 molecular cloud (see Figures~\ref{uf2}a and~\ref{uf2}b). 
In the velocity space, there is only one velocity component observed in the direction of the EFS 
(see Figures~\ref{ufg2}b and~\ref{ufg2}d). 
This implies the existence of a single EFS. 
The peak positions of ten dust continuum sources at 1.1 mm are also shown in Figures~\ref{uf2}a and~\ref{uf2}b and 
are mainly distributed toward the EFS (also see Section~\ref{subsec:temp} for quantitative estimate). 
This particular result gives a hint about fragmentation of the filamentary cloud. 
To further infer this signature, we estimated virial mass (M$_{vir}$) and virial parameter (M$_{vir}$/$M_{cloud}$) of the S242 molecular cloud using the observed physical parameters.  
Using the NANTEN $^{13}$CO (1-0) line data (beam size $\sim$2$'$.7), \citet{kawamura98} reported M$_{cloud}$ ($\sim$7000 M$_{\odot}$), radius (R$_{c}$ $\sim$7 pc), and line width ($\Delta V$ = 2.1 km s$^{-1}$) for the S242 molecular cloud. 
The virial mass of a cloud of radius R$_{c}$ (in pc) and line-width $\Delta V$ (in km s$^{-1}$) is defined 
as M$_{vir}$ ($M_\odot$)\,=\,k\,R$_{c}$\,$\Delta V^2$ \citep{maclaren88}, 
where the geometrical parameter, k\,=\,126, for a density profile $\rho$ $\propto$ 1/r$^2$. 
A virial parameter less than 1 indicates the cloud prone to collapse, and greater than 1 is resistant to collapse. 
In the present case, we obtain M$_{vir}$ $\sim$3890 $M_\odot$, which is less than M$_{cloud}$. 
This implies that the virial parameter is less than 1, suggesting the cloud is unstable against gravitational collapse.

In Figures~\ref{uf2}a and~\ref{uf2}b, we find that the S242 site has a shell-like appearance where noticeable warm dust emission at MIR 
is seen. 
In general, the ionized gas and the warm dust emissions are seen systematically correlated within H\,{\sc ii} regions \citep[e.g.][]{deharveng10}. 
With the knowledge of the presence and absence of the radio continuum emission, we find two distinct ends of the EFS. 
One EFS end contains the S242 H\,{\sc ii} region, while the other EFS end is seen without noticeable ionized emission (see Figure~\ref{uf2}b). 
Note that a majority of the dust continuum sources at 1.1 mm are seen at both the ends of the EFS. 

Together, Figures~\ref{uf1} and~\ref{uf2} allow to probe the S242 molecular cloud, EFS, dust continuum sources at 1.1 mm, and H\,{\sc ii} region in our selected 
site probed in this paper.
\subsubsection{{\it Herschel} temperature and column density maps}
\label{subsec:temp}
\begin{figure*}
\epsscale{1}
\plotone{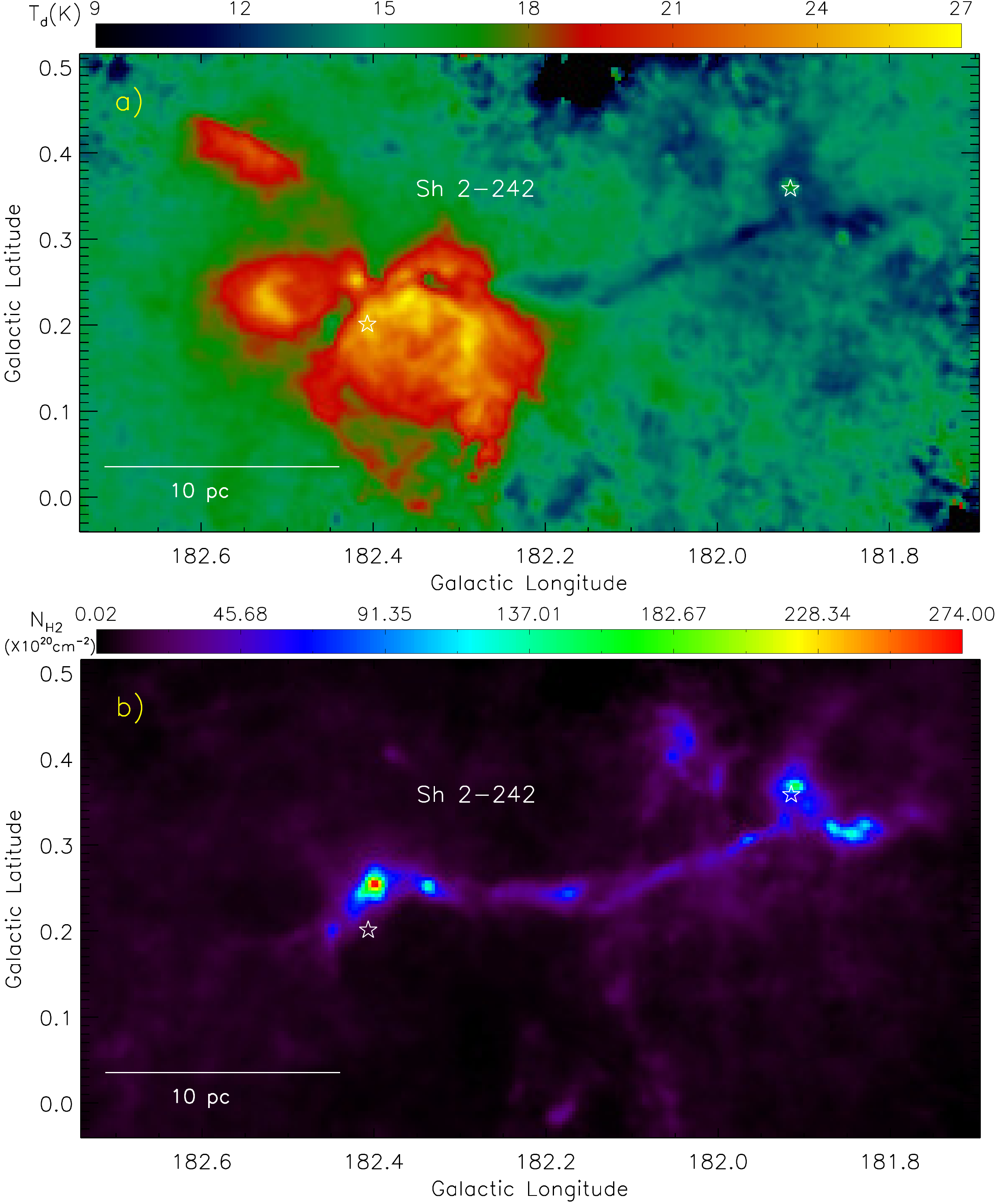}
\caption{\scriptsize a) {\it Herschel} temperature map of the region around S242 (see text for details).
b) {\it Herschel} column density ($N(\mathrm H_2)$) map of the region around S242 (see text for details). 
The column density map can be used to infer the extinction using the relation $A_V=1.07 \times 10^{-21}~N(\mathrm H_2)$ \citep{bohlin78} 
and to identify the clumps (see text for details). In both the panels, other marked symbols and labels are similar to those shown in Figure~\ref{uf1}.}
\label{uf3}
\end{figure*}

In this section, the temperature and column density maps derived using the {\it Herschel} continuum images are discussed. 
The final temperature and column density maps (resolution $\sim$37$\arcsec$) are shown in Figures~\ref{uf3}a and~\ref{uf3}b, respectively.

\begin{figure*}
\epsscale{0.8}
\plotone{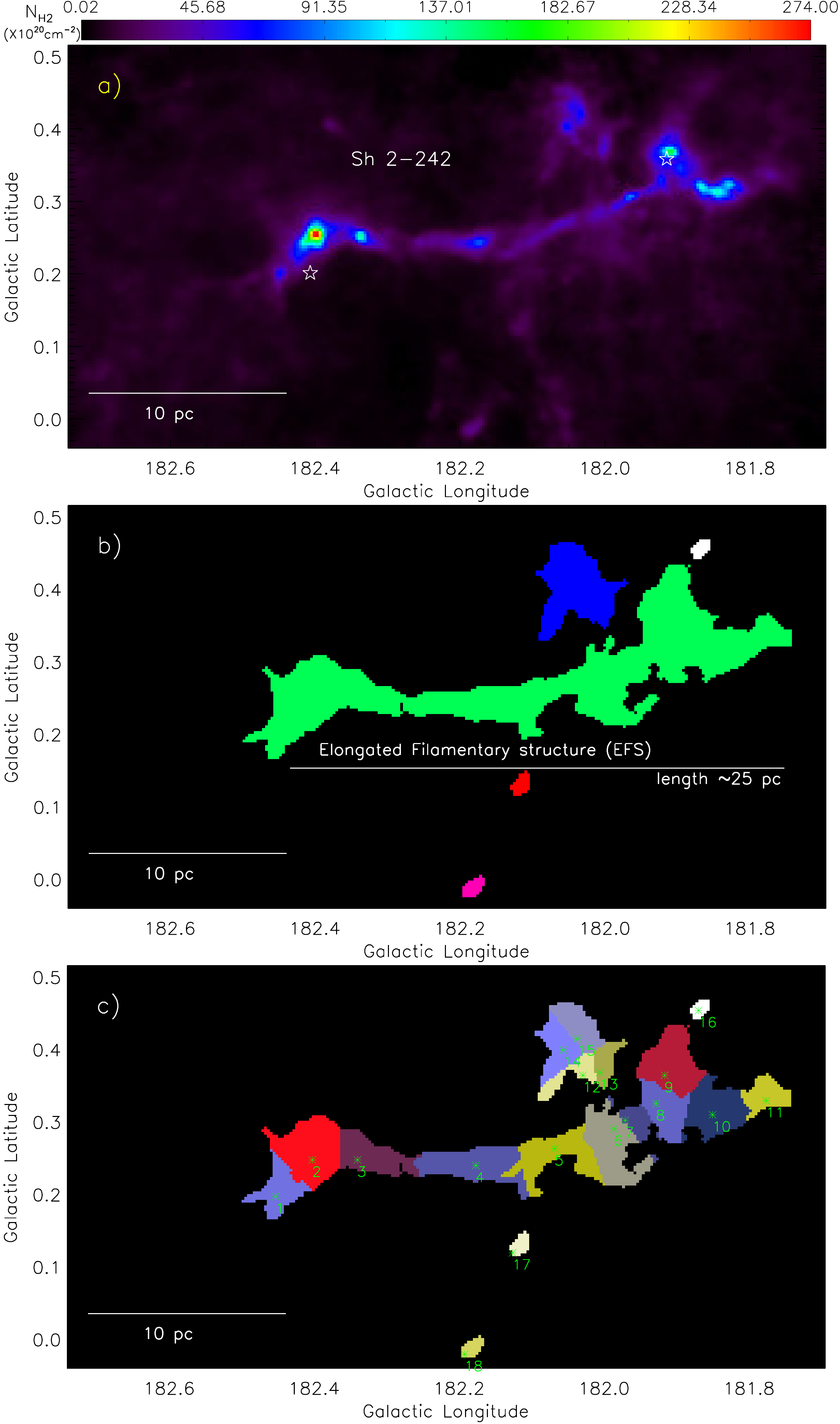}
\caption{\scriptsize a) {\it Herschel} column density ($N(\mathrm H_2)$) map of the region around S242.
b) An elongated filamentary structure (EFS) is traced in the column density map at a contour level of 1.5 $\times$ 10$^{21}$ cm$^{-2}$.
c) The identified clumps are highlighted by asterisks and the boundary of each {\it Herschel} clump is also 
shown in figure. The boundary of each {\it Herschel} clump is highlighted along with its corresponding clump ID.}
\label{uf4}
\end{figure*}

In the {\it Herschel} temperature map, the S242 H\,{\sc ii} region is associated with the considerably warmer gas (T$_{d}$ $\sim$22-32 K)
(see Figure~\ref{uf3}a). The map clearly traces the spatial extent of warm dust emission linked with the S242 site, 
where the ionized emission is observed. The temperature map reveals temperature variations toward the EFS (see areas near both the EFS ends). 
The EFS is traced in a temperature range of about 10--12~K away from the S242 H\,{\sc ii} region. In Figures~\ref{uf3}b,~\ref{uf4}a, and~\ref{uf4}b, the EFS (length $\sim$25 pc) is evident in 
the {\it Herschel} column density map at a contour level of 1.5 $\times$ 10$^{21}$ cm$^{-2}$ 
and several condensations are also seen toward this feature (also see Figure~\ref{uf4}c). 
The S242 site is located in the highest column density region (peak $N(\mathrm H_2)$ $\sim$2.7~$\times$~10$^{22}$ cm$^{-2}$; A$_{V}$ $\sim$29 mag). 
Here, we used a relation \citep[$A_V=1.07 \times 10^{-21}~N(\mathrm H_2)$;][]{bohlin78} between optical extinction and hydrogen column density. 
In the column density map (see Figure~\ref{uf4}a), the ``{\it clumpfind}" IDL program \citep{williams94} is employed to trace the clumps and 
to find their total column densities. 
We used several column density contour levels as an input parameter for the ``{\it clumpfind}" and the lowest contour level was assigned at 3$\sigma$.
Eighteen clumps are identified in the map and are labeled in Figure~\ref{uf4}c. 
Furthermore, the boundary of each clump is also shown in Figure~\ref{uf4}c. 
Eleven out of eighteen clumps (e.g., 1 to 11) are found toward the EFS. 
We have also determined the mass of each clump using its total column density. 
The mass of a single {\it Herschel} clump is estimated using the following formula:
\begin{equation}
M_{clump} = \mu_{H_2} m_H Area_{pix} \Sigma N(H_2)
\end{equation}
where $\mu_{H_2}$ is assumed to be 2.8, $Area_{pix}$ is the area subtended by one pixel, and 
$\Sigma N(\mathrm H_2)$ is the total column density. 
The mass of each {\it Herschel} clump is tabulated in Table~\ref{tab1}. 
The table also lists effective radius, peak column density, peak temperature, and mean central number density (n$_{c}$) of each clump.
The clump masses vary between 25 M$_{\odot}$ and 1020 M$_{\odot}$. 
We also find peak temperatures, mean central number densities, and peak column densities (corresponding extinction) of the clumps 
ranging from 10--26 K, 505--2575 cm$^{-3}$, and 1.9--27 $\times$ 10$^{21}$ cm$^{-2}$ (A$_{V}$ = 2-29 mag), respectively.
The mean central number density of each clump refers to the average number density along the line of sight and is obtained from the 
peak column density divided by the size of each clump. Eleven clumps, which are distributed toward the EFS, have masses varying between 150 and 1020 M$_{\odot}$. 
Interestingly, massive clumps (M$_{clump}$ $\sim$260, 700, 700, and 1020 M$_{\odot}$) are spatially seen at both the EFS ends (also see Table~\ref{tab1}).
The virial mass analysis of these clumps is not performed in this paper, due to to non-availability of optically thin line data (such as, NH$_{3}$ and CS).

We have also obtained a total column density inside the contour of $N(\mathrm H_2)$ = 1.5 $\times$ 10$^{21}$ cm$^{-2}$ and 
have computed a total mass of the EFS (M$_{EFS}$) to be $\sim$5000 M$_{\odot}$. 
Adopting the value of M$_{EFS}$ and length of EFS ($\sim$25 pc), the mass per unit length is calculated to be $\sim$200 M$_{\odot}$ pc$^{-1}$. 
Note that there is no knowledge of the inclination angle, i, of the EFS, and for reference, we have adopted here i = 0.
Due to the inclination, the line mass can be affected by a factor of cos i \citep[e.g.,][]{kainulainen16}. 
Hence, the observed mass per unit length value can be considered as an upper limit.
\subsection{Star formation activities in and around S242}
\label{subsec:phot1}
The infrared excess emission displayed by sources is an extremely powerful utility to probe the embedded young stellar populations. 
In a given star-forming region, the knowledge of spatial distribution of these young stellar populations helps to infer the star formation activities.
In this section, we describe the identification and classification schemes of young stellar objects (YSOs) using the GPS-2MASS and 
GLIMPSE360 photometric data from 1--5 $\mu$m. 
Furthermore, to investigate the young stellar clusters, the distribution of YSOs is also presented in and around S242.\\

\begin{figure*}
\epsscale{0.85}
\plotone{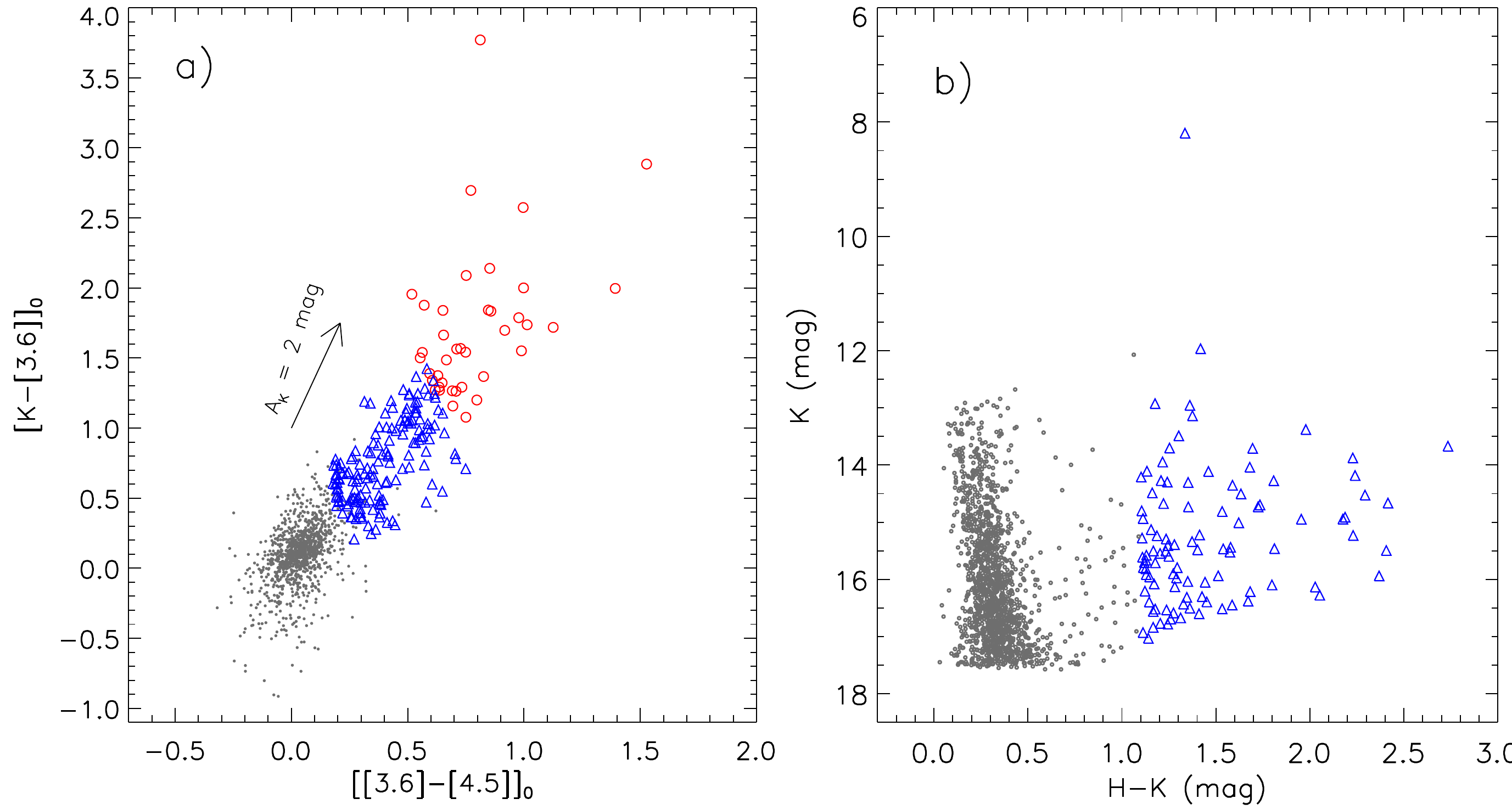}
\epsscale{0.85}
\plotone{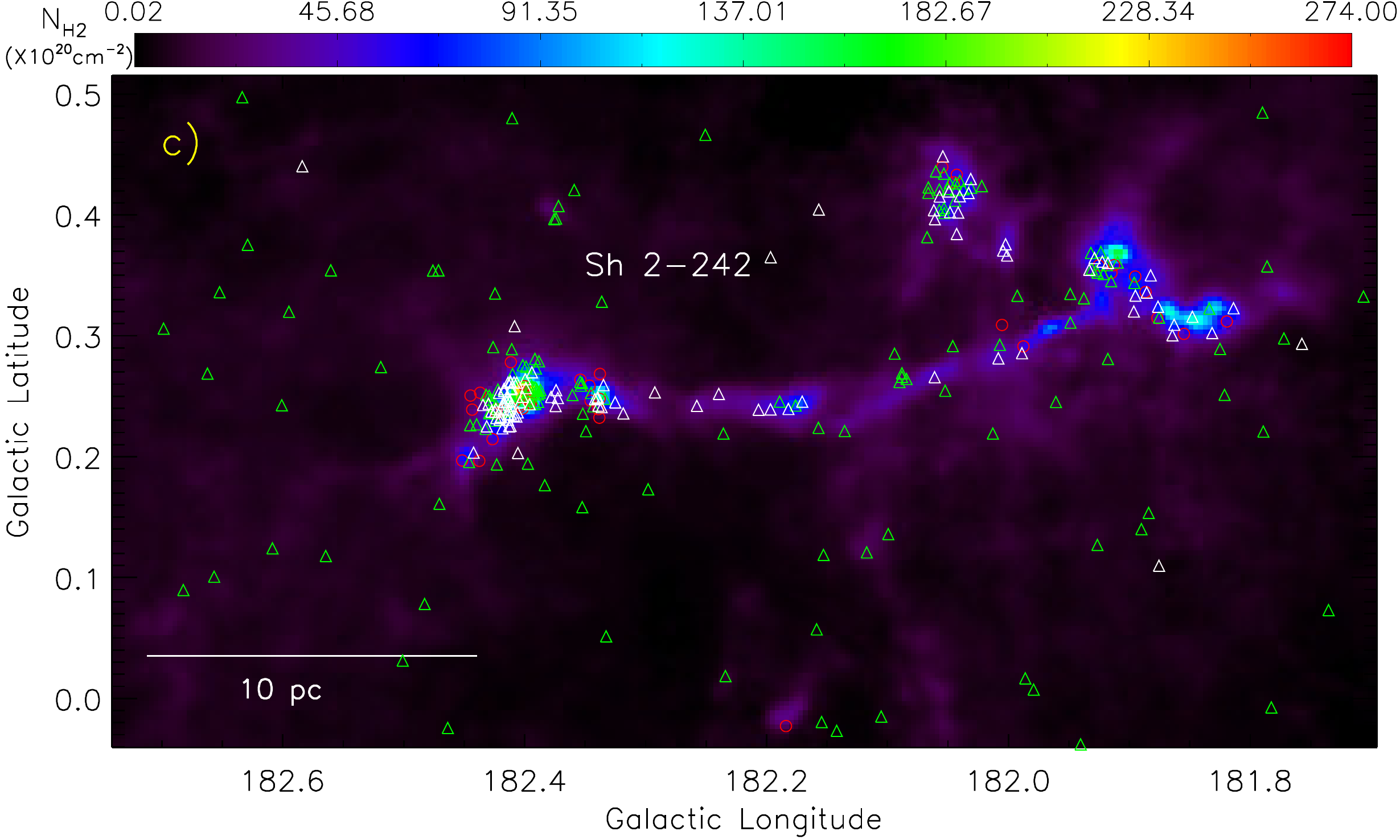}
\caption{\scriptsize The identification of young stellar populations within the region around S242 (see Figure~\ref{uf2}a). 
a) The dereddened [K$-$[3.6]]$_{0}$ $vs$ [[3.6]$-$[4.5]]$_{0}$ color-color diagram using 
the H, K, 3.6 $\mu$m, and 4.5 $\mu$m data (see text for details). 
The extinction vector is shown using the average extinction laws from \citet{flaherty07}. 
b) Color-magnitude diagram (H$-$K/K) of the sources detected only in H and K bands that have no counterparts 
in our selected GLIMPSE360 catalog.
In each panel, Class~I and Class~II YSOs are marked by red circles and open blue triangles, respectively, 
while, the dots (in gray) refer the stars with only photospheric emissions. 
In the color-color diagram, we have shown only 1101 out of 8257 stars with photospheric emissions. 
In the color-magnitude diagram, we have shown only 1501 out of 20998 stars with photospheric emissions. 
Due to large numbers of stars with photospheric emissions, we have randomly shown only some of these stars in the diagrams. 
The positions of all the identified YSOs are shown in Figure~\ref{uf5}c.
c) The spatial distribution of selected YSOs within the region around S242. 
The positions of YSOs are overlaid on the {\it Herschel} column density map. 
The background map is similar to the one shown in Figure~\ref{uf4}a. 
The positions of Class~I and Class~II YSOs are shown by circles and triangles, respectively. 
The YSOs selected using the H, K, 3.6 $\mu$m, and 4.5 $\mu$m data (see Figure~\ref{uf5}a) 
are shown by red circles and green triangles, whereas the white triangles represent the YSOs 
identified using the H and K bands (see Figure~\ref{uf5}b).}
\label{uf5}
\end{figure*}

The dereddened color-color space ([K$-$[3.6]]$_{0}$ and [[3.6]$-$[4.5]]$_{0}$) is a very promising 
tool to identify infrared-excess sources \citep[e.g.][]{gutermuth09}. 
We computed the dereddened color-color plot ([K$-$[3.6]]$_{0}$ and [[3.6]$-$[4.5]]$_{0}$) 
using the GLIMPSE360 and 2MASS photometric data at 1--5 $\mu$m. 
The dereddened colors were obtained using the color excess ratios listed in \citet{flaherty07}. 
Using the dereddened color conditions presented in \citet{gutermuth09}, 
we obtain 192 (39 Class~I and 153 Class~II) YSOs in our selected region probed in this paper. 
One can also infer possible dim extragalactic contaminants from the selected YSOs with additional 
conditions (i.e., [3.6]$_{0}$ $<$ 15 mag for Class~I and [3.6]$_{0}$ $<$ 14.5 mag for Class~II) \citep[e.g.,][]{gutermuth09}.  
The dereddened 3.6 $\mu$m magnitudes were obtained using the observed color and the reddening laws \citep[from][]{flaherty07}.
In Figure~\ref{uf5}a, we show the [K$-$[3.6]]$_{0}$ versus [[3.6]$-$[4.5]]$_{0}$ color-color plot for the sources. 
The selected Class~I and Class~II YSOs are highlighted by the red circles and blue triangles, respectively.\\  
We also find some additional YSOs having detections only in the H and K bands using the GPS-2MASS data. 
Note that the UKIDSS-GPS NIR data have better spatial resolution and are three magnitudes deeper than those of 2MASS data. 
We employed a color-magnitude (H$-$K/K) diagram to obtain additional young stellar populations, which are shown in Figure~\ref{uf5}b.
The diagram allows to depict the red sources having H$-$K $>$ 1.1 mag. 
We obtained this color criterion based on the color-magnitude analysis of a nearby control field. 
In Figure~\ref{uf5}b, the selected young stellar populations are highlighted by blue triangles. 
We identify 101 additional YSOs using the color-magnitude space in our selected region.\\ 
Using the GPS-2MASS and GLIMPSE360 data at 1--5 $\mu$m, our analysis yields a total of 293 YSOs 
in our selected region around S242. In Figure~\ref{uf5}c, we have overlaid the positions of all 
the selected YSOs on the {\it Herschel} column density map. A majority of these YSOs are distributed toward the 
filamentary structure. 
The Class~I YSOs are located only in the areas of high column density.
\subsection{Clustering of YSOs}
\label{subsec:phot2}
\begin{figure*}
\epsscale{0.88}
\plotone{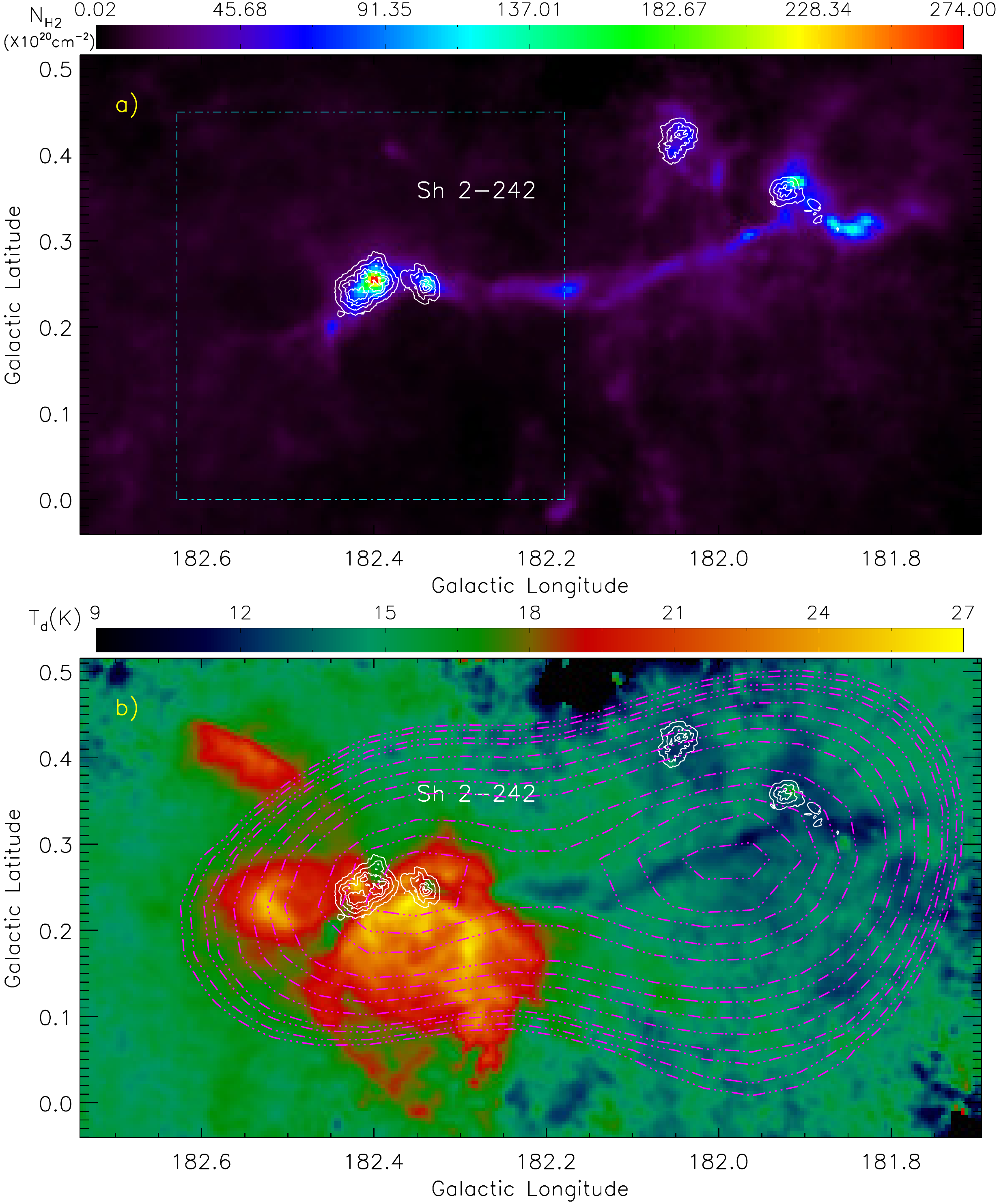}
\caption{\scriptsize a) The {\it Herschel} column density map is overlaid with the surface density contours (in white) of 
the identified YSOs toward the EFS containing the S242 site. 
The contours are shown at 5, 10, 20, and 40 YSOs/pc$^{2}$, from the outer to the inner side. 
The dotted-dashed cyan box encompasses the area shown in Figures~\ref{uf7}a and~\ref{uf7}b. 
b) The {\it Herschel} temperature map is overlaid with the surface density contours (in white) of 
the identified YSOs toward the EFS containing the S242 site. 
The $^{12}$CO emission contours (dotted-dashed magenta) are also superimposed with levels of 43, 45, 48, 50, 55, 60, 65, 70, 80, 85, 90, 94, and 98 \% of 
the peak value (i.e., 10.478 K km s$^{-1}$). 
The integrated $^{12}$CO intensity map reveals a continuous velocity structure in the direction of S242. 
The $^{12}$CO map is smoothed with a Gaussian function with radius of three.} 
\label{uf6}
\end{figure*}

In this section, we study the individual groups or clusters of YSOs based on their spatial distribution and the statistical surface density utility.
In Figures~\ref{uf6}a and~\ref{uf6}b, to trace the groups/clusters of YSOs, we have superimposed the surface density contours of YSOs on the {\it Herschel} column density and temperature maps. 
The surface density map of YSOs is produced using the nearest-neighbour (NN) technique \citep[e.g.,][]{gutermuth09,bressert10,dewangan15}. 
Using a 5$\arcsec$ grid and 6 NN at a distance of 2.1 kpc, the surface density map of 293 YSOs is computed in a manner similar to that described in \citet{dewangan15}. 
The clusters of YSOs are mainly seen at both the EFS ends (see Figures~\ref{uf6}a and~\ref{uf6}b), however, there is a noticeable presence of young stellar populations toward the EFS without any clustering away from its both the ends (i.e. its center; see Figure~\ref{uf5}c). 
Furthermore, a cluster of YSOs is also evident toward the {\it Herschel} clumps, nos. 14 and 15, which are spatially distributed away from the EFS (see Figure~\ref{uf6}a).

Together, the star formation activities have been found toward the clumps linked with the EFS and other {\it Herschel} clumps.
\subsubsection{S242 H\,{\sc ii} region}
\label{subsec:hii}
\begin{figure*}
\epsscale{0.65}
\plotone{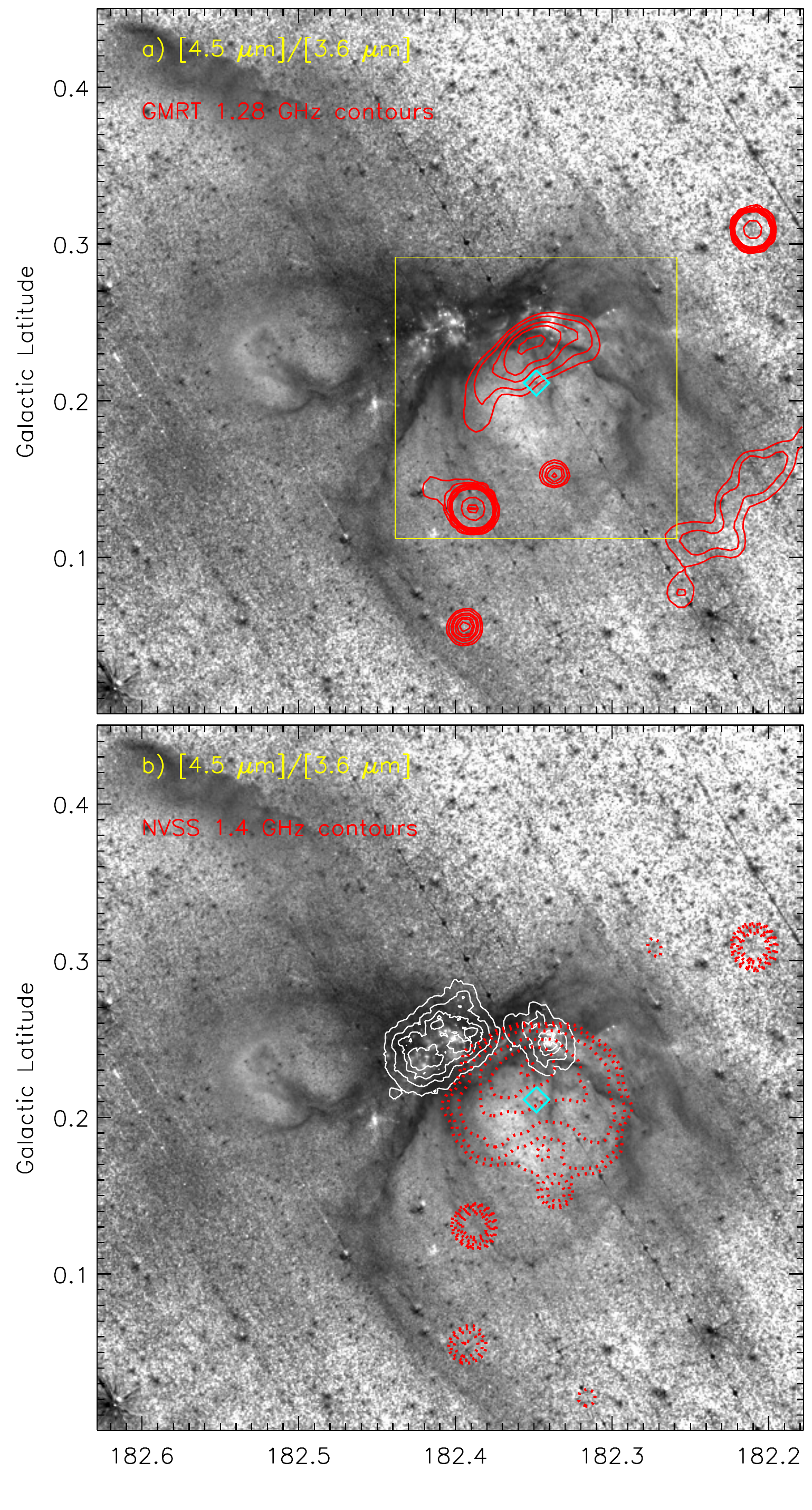}
\caption{\scriptsize a) {\it Spitzer} ratio map of 4.5 $\mu$m/3.6 $\mu$m emission is superimposed with the GMRT radio continuum 
emission at 1.28 GHz (beam size $\sim$20\farcs4 $\times$ 20\farcs4). 
The radio contours are shown with levels of 5, 7, 10, 12, 15, 17, 20, 60, 90, and 98\% of the peak value (i.e. 13.790 mJy/beam). 
The solid yellow box encompasses the area shown in Figures~\ref{uf8}a,~\ref{uf8}b,~\ref{uf8}c,~\ref{uf8}d, and~\ref{uf9}. 
b) {\it Spitzer} ratio map of 4.5 $\mu$m/3.6 $\mu$m emission is overlaid with the NVSS radio continuum emission at 1.4 GHz. 
The NVSS 1.4 GHz dotted contours (in red) are shown with levels of 3.5, 5.5, 10, 20, 30, and 40\% of the peak value (i.e. 0.377 Jy/beam). 
Figure also shows the surface density contours (in white) of the identified YSOs. 
The contours are shown at 5, 10, 20, and 40 YSOs/pc$^{2}$, from the outer to the inner side. In each panel, a diamond symbol (in cyan) 
indicates the location of a massive star, BD+26 980. In both the panels, the ratio map is subjected to median filtering with a width 
of 4 pixels and smoothened by 4 pixel $\times$ 4 pixel using a boxcar algorithm.}
\label{uf7}
\end{figure*}

In Figure~\ref{uf7}, we present a zoomed-in view of the S242 site using the {\it Spitzer}-IRAC ratio map and the radio continuum maps.
In Figures~\ref{uf7}a and~\ref{uf7}b, the area around the S242 site is chosen based on the spatial extent of the warm dust emission (see Figure~\ref{uf6}b). 
In combination with the radio continuum map, the {\it Spitzer}-IRAC ratio map of 4.5 $\mu$m/3.6 $\mu$m emission is used here to infer the signatures of molecular 
outflows and the impact of a massive star on its surroundings \citep[e.g.,][]{dewangan16,dewangan17a}. 
IRAC 3.6 $\mu$m band harbors polycyclic aromatic hydrocarbon (PAH) emission 
at 3.3 $\mu$m as well as a prominent molecular hydrogen feature at 3.234 $\mu$m ($\nu$ = 1--0 $ O$(5)). 
IRAC 4.5 $\mu$m band contains a hydrogen recombination line Br$\alpha$ (4.05 $\mu$m) and a prominent molecular hydrogen line emission ($\nu$ = 0--0 $S$(9); 4.693 $\mu$m), which is produced by outflow shocks. It is known that IRAC 3.6 $\mu$m and 4.5 $\mu$m images have 
almost identical point response functions, therefore the ratio of 4.5 $\mu$m to 3.6 $\mu$m images can be used to directly produce a ratio map of 4.5 $\mu$m/3.6 $\mu$m emission (see Figure~\ref{uf7}) \citep[e.g.,][]{dewangan16}. 
In Figure~\ref{uf7}a, we infer the bright and dark/black regions in the ratio map of 4.5 $\mu$m/3.6 $\mu$m emission. 
In ratio 4.5 $\mu$m/3.6 $\mu$m map, the bright emission regions trace the excess of 4.5 $\mu$m emission, 
while the black or dark gray regions depict the excess of 3.6 $\mu$m emission. 
In Figures~\ref{uf7}a and~\ref{uf7}b, the ionized emissions at 1.28 and 1.4 GHz are distributed within a shell-like morphology. 
Due to the presence of 3.3 $\mu$m PAH feature in the 3.6 $\mu$m band, the dark/black regions surrounding the ionized emission appear to trace 
photodissociation regions (or photon-dominated regions, or PDRs). Furthermore, the bright emission regions at one of the ends of the EFS 
containing the S242 site, where the radio continuum emission is absent and three clusters of YSOs are found, appear 
to probably trace the outflow activities (see Figure~\ref{uf7}b). 
This interpretation can be supported by the presence of the CO and H$_{2}$ outflow signatures in the embedded clusters near IRAS 05490+268
\citep[e.g.,][]{snell90,varricatt10}. 

\begin{figure*}
\epsscale{1}
\plotone{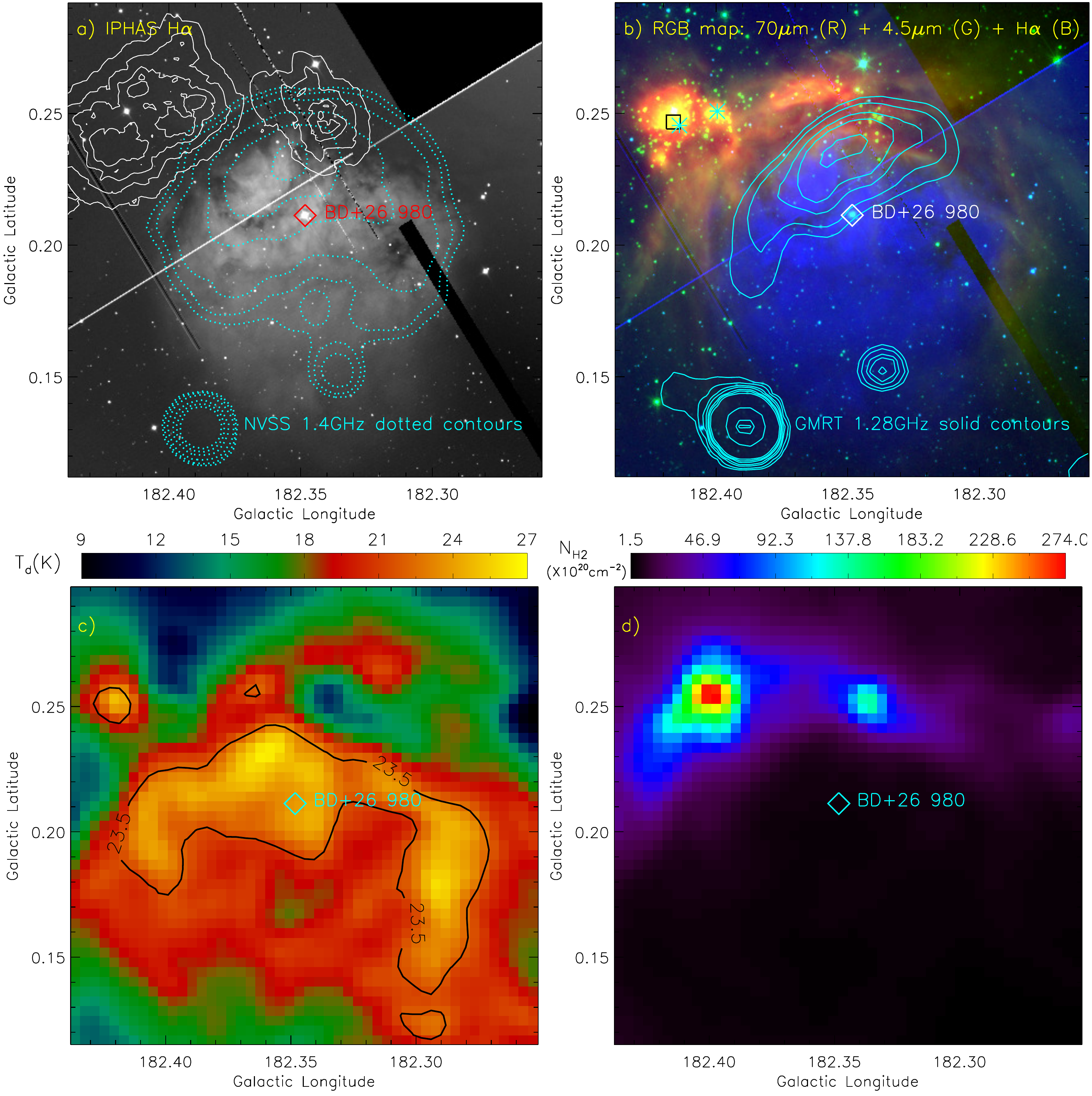}
\caption{\scriptsize Zoomed-in veiw of the S242 site. a) The IPHAS H$\alpha$ gray-scale image is superimposed with the NVSS 1.4 GHz contours. 
The NVSS 1.4 GHz dotted contours (in cyan) are shown with levels of 4, 6, 10, 20, 30, and 40\% of the 
peak value (i.e. 0.377 Jy/beam). Figure also shows the surface density contours (in white) of the identified YSOs. 
The contours are shown at 5, 10, 20, and 40 YSOs/pc$^{2}$, from the outer to the inner side. 
b) Color composite map is the result of the combination of 70 $\mu$m (red), 4.5 $\mu$m (green), and H$\alpha$ (blue)) 
and is superimposed with the GMRT 1.28 GHz contours. A square indicates the location of IRAS 05490+2658. 
Asterisk symbols (in cyan) represent the locations of two 1.2 mm dust continuum peaks \citep[from][]{beuther02}. 
c) {\it Herschel} temperature map of the S242 site is shown with a temperature contour level of 23.5 K. 
d) {\it Herschel} column density map of the S242 site. In all the panels, a diamond symbol indicates the location of BD+26 980.}
\label{uf8}
\end{figure*}

In Figure~\ref{uf8}a, we present an H$\alpha$ image overlaid with the NVSS 1.4 GHz emission, 
showing the spatial match between the radio continuum emission and the H$\alpha$ emission. 
The position of a previously characterized B0V star (BD +26 980) appears near the peak of radio continuum emission. 
The surface density contours of the identified YSOs are also highlighted in Figure~\ref{uf8}a. 
In Figure~\ref{uf8}b, we show a three-color composite map made using the {\it Herschel} 70 $\mu$m in red, {\it Spitzer} 4.5 $\mu$m in green, and 
IPHAS H$\alpha$ in blue. The color composite map is superimposed with the GMRT 1.28 GHz emission and positions of two 1.2 mm dust continuum 
peaks \citep[from][]{beuther02}. In order to compare the spatial distribution of dust temperature and column density with the ionized emission, the {\it Herschel} temperature and column density maps of the S242 site are shown in Figures~\ref{uf8}c and~\ref{uf8}d, respectively. 
We find the YSO clusters toward the high column density materials. Additionally, there is a lack of column density material within the S242 H\,{\sc ii} region. 
Earlier in Section~\ref{subsec:temp}, we mentioned the spatial correlation between the warm dust emission and the ionized emission (also see Figures~\ref{uf8}b and~\ref{uf8}c). In Figure~\ref{uf8}, one can infer the zoomed-in view of the S242 site, tracing the spatial location of the ionized emission, warm dust emission, high column density material, and the embedded stellar populations.

Using the radio continuum maps, we also compute the Lyman continuum photons and the 
dynamical age (t$_{dyn}$) of the S242 H\,{\sc ii} region. 
Using the NVSS 1.4 GHz map and the {\it clumpfind} algorithm, the integrated flux density (S$_{\nu}$) 
and the radius (R$_{HII}$) of the H\,{\sc ii} region are determined to be 437 mJy and 1.84 pc, respectively. 
Following the equation given in \citet{matsakis76}, the number of Lyman continuum photons (N$_{uv}$) is computed 
to be $\sim$1.5 $\times$ 10$^{47}$ s$^{-1}$ (logN$_{uv}$ $\sim$47.18) for the S242 H\,{\sc ii} region \citep[see][for more details]{dewangan16}. 
In this analysis, we used the integrated flux density value, a distance of 2.1 kpc, and the electron temperature of 10000~K. 
The value of N$_{uv}$ is found to be consistent with a single ionizing star of spectral type B0.5--B0V (see Table II in \citet{panagia73} and also \citet{smith02}).  
The estimate of N$_{uv}$ at GMRT 1.28 GHz frequency also corresponds to a single ionizing star of B0.5--B0V spectral type.
These calculations are also in agreement with the previously reported spectral type of the ionizing source of the S242 site \citep{hunter90}.  

The equation of the dynamical age of the H\,{\sc ii} region is given below at a radius R$_{HII}$ \citep[e.g.,][]{dyson80}:
\begin{equation}
t_{dyn} = \left(\frac{4\,R_{s}}{7\,c_{s}}\right) \,\left[\left(\frac{R_{HII}}{R_{s}}\right)^{7/4}- 1\right] 
\end{equation}
where c$_{s}$ is the isothermal sound velocity in the ionized gas (c$_{s}$ = 11 km s$^{-1}$; \citet{bisbas09}), 
R$_{HII}$ is mentioned above, and R$_{s}$ is the radius of the Str\"{o}mgren sphere (= (3 N$_{uv}$/4$\pi n^2_{\rm{0}} \alpha_{B}$)$^{1/3}$, where 
the radiative recombination coefficient $\alpha_{B}$ =  2.6 $\times$ 10$^{-13}$ (10$^{4}$ K/T)$^{0.7}$ cm$^{3}$ s$^{-1}$ \citep{kwan97}, 
N$_{uv}$ is mentioned above, and ``n$_{0}$'' is the initial particle number density of the ambient neutral gas). 
Considering a typical value of n$_{0}$ (=10$^{3}$ cm$^{-3}$), we obtained the dynamical age of the S242 H\,{\sc ii} region 
to be $\sim$0.5 Myr. 

\begin{figure*}
\epsscale{1}
\plotone{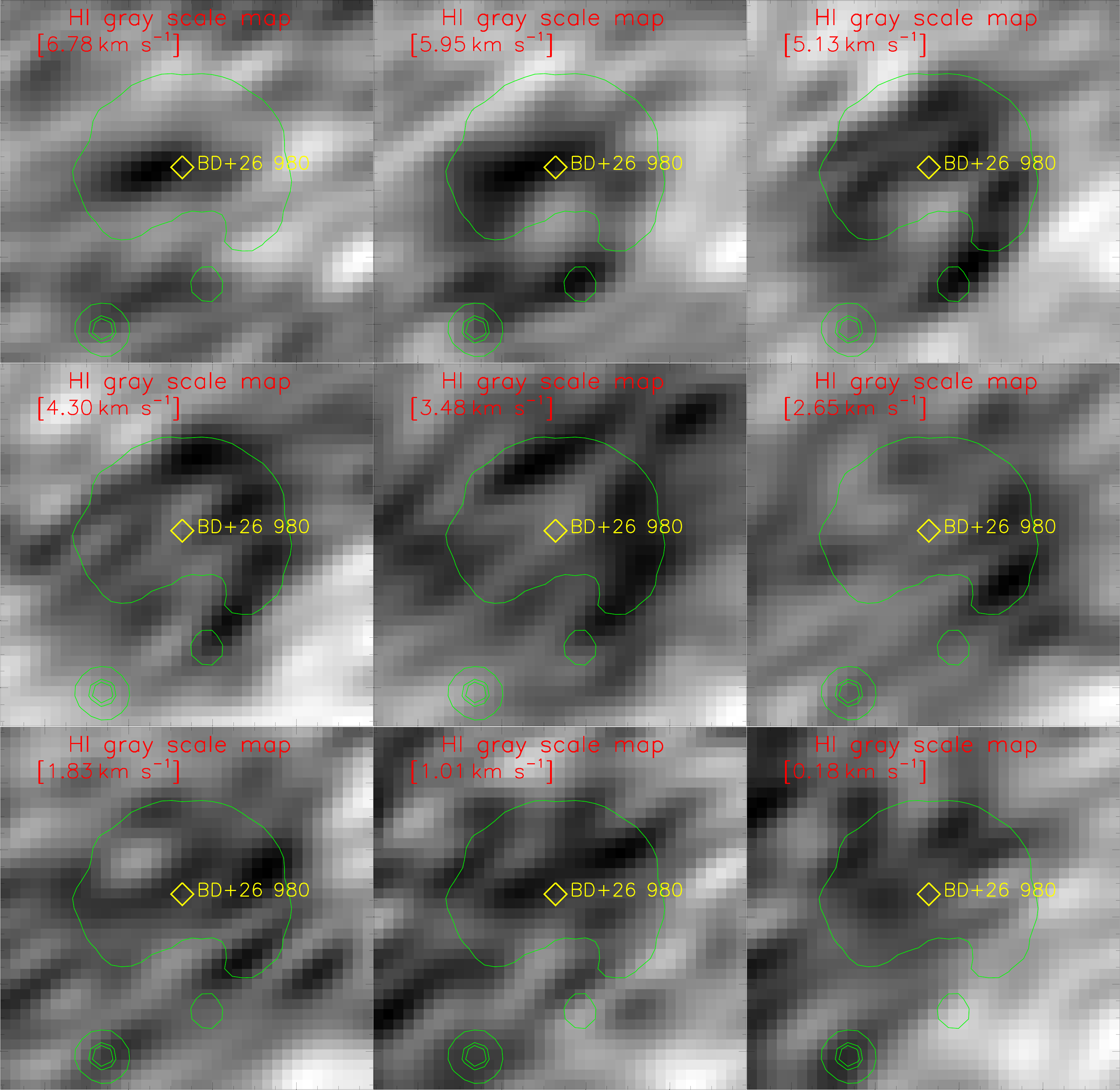}
\caption{\scriptsize The CGPS 21 cm H\,{\sc i} velocity channel maps of the S242 site. 
The velocity value (in km s$^{-1}$) is labeled in each panel. 
In each panel, the NVSS 1.4 GHz contours (in green) are superimposed with levels of 8, 55, and 70\% of 
the peak value (i.e., 0.377 Jy/beam), tracing the ionized hydrogen emission. The location of a massive star BD+26 980 is also marked in the maps. 
The sphere-like shell morphology of the S242 site is depicted as H\,{\sc i} self-absorption features 
\citep[i.e., cold H\,{\sc i} observed in absorption against warmer background H\,{\sc i};][]{kerton05}.}
\label{uf9}
\end{figure*}

We have also utilized 21 cm H\,{\sc i} line data toward the S242 site. 
Figure~\ref{uf9} shows 21 cm H\,{\sc i} velocity channel maps of the S242 site. 
We find the black or dark gray regions in the H\,{\sc i} channel maps which trace the H\,{\sc i} self-absorption (HISA) features \citep[e.g.,][]{kerton05}. 
In a velocity range of 3.48 to 5.95 km s$^{-1}$, the shell-like HISA feature is evident in the channel maps. 
The existence of the HISA features can be explained by the presence of the residual amounts of very cold H\,{\sc i} gas in molecular clouds 
\citep{burton78,baker79, burton81,liszt81,dewangan17a}. 
\section{Discussion}
\label{sec:disc}
The cold dust continuum emission traced in the 160--1100 $\mu$m images has been used to probe the elongated filaments and the distribution of clumps in these filaments \citep[e.g.][and references therein]{schneider12,ragan14,contreras16,li16}. It has been observed that the large-scale filaments are found to be unstable to radial collapse and fragmentation. One of the key questions in star formation research is how the filaments fragment into dense clumps/cores that form star.
The gravitational fragmentation process can be employed to explain the presence of ``dense clumps/cores" in filaments and can be inferred with a knowledge of the line mass of the filament ($M_{\rm line}$) \citep[e.g.,][]{andre10}. 
It has been suggested that thermally supercritical filaments are associated with the prestellar clumps/cores and star formation activity, where the mass per unit length greater than the critical 
value ($M_{\rm line,crit}$) (i.e., $M_{\rm line} > M_{\rm line,crit}$). On the other hand, thermally subcritical filaments ($M_{\rm line} < M_{\rm line,crit} $) 
often lack {\it Herschel} 
prestellar clumps/cores and embedded protostars \citep{andre10}. 
The critical mass per unit length ($M_{\rm line,crit}~=~2 c^{2}_{\rm s}/G$; where $c_{\rm s}$ is the isothermal sound speed (i.e. $\sim$0.2~km s$^{-1}$ at T = 10 K) and $G$ is the gravitational constant) is needed for a filament to be gravitationally unstable to radial contraction and fragmentation along its length \citep{inutsuka97}. 
A critical line mass $M_{\rm line,crit} $ is equal to $\sim$16~M$_{\odot}$/pc $\times$ (T$_{gas}$/10 K) for gas filaments \citep{andre14}, 
hence one can find $M_{\rm line,crit}$ $\sim$16~M$_{\odot}$/pc at T= 10~K  \citep[e.g.,][]{ostriker64,kainulainen16}.

Using the $^{13}$CO line data analysis, the elongated S242 molecular cloud is unstable against gravitational collapse. 
The EFS embedded within the S242 molecular cloud is the most prominent feature observed in the {\it Herschel} column density map. 
The {\it Herschel} temperature map traces $\sim$10-12 K toward the EFS (except close to the S242 H\,{\sc ii} region). 
The observed mass per unit length of EFS is computed to be 
$\sim$200 M$_{\odot}$ pc$^{-1}$, which is much higher than the critical masses per unit length (16--48 M$_{\odot}$/pc) at T = 10--30 K. 
For a comparison purpose, \citet{andre16} presented a Table~1 containing the physical parameters of some well-documented filaments 
($M_{\rm line}$ $\sim$4500 M$_{\odot}$ pc$^{-1}$ (DR21), $\sim$290 M$_{\odot}$ pc$^{-1}$ (Serpens South), $\sim$50 M$_{\odot}$ pc$^{-1}$ (Taurus B211/B213), and $\sim$20 M$_{\odot}$ pc$^{-1}$ (Musca)). 
The EFS also contains a chain of {\it Herschel} clumps (M$_{clump}$ $\sim$150 to 1020 M$_{\odot}$), depicting the evidence of fragmentation along its length. 
The most massive clumps are seen at both the EFS ends and the S242 H\,{\sc ii} region is traced at one of the EFS ends. 
The young stellar populations are also spatially found toward the EFS, indicating a convincing evidence of ongoing star formation 
activities (see Section~\ref{subsec:phot1}). However, the clusters of YSOs are only observed at both the EFS ends (see Section~\ref{subsec:phot2}). 
Most recently, \citet{kainulainen16} also found the fragmentation strongly at the ends of the Musca cloud. 
In the long (but finite-sized) filaments, numerical and analytical studies showed that the acceleration of gas is a function of the relative position along the filament and is the greatest at its both the ends \citep[][see also \citealt{burkert04} for a study in a sheet geometry]{bastien83, pon11, pon12, clarke15}. These higher accelerations indicate shorter local collapse time-scales, suggesting the formation of fragments at the ends of the filament prior to its center \citep{pon11, pon12}. 
\citet{pon11} reported that the local collapse may act a factor of two-to-three faster at the ends of the filament than at its center \citep[see Figures~5 and~6 in][]{pon11}. Additionally, \citet{heitsch08} presented three-dimensional models of molecular cloud formation in large-scale colliding flows including self-gravity, 
and found that global collapse of a molecular cloud produces centrally located large-scale filaments, while local gravitational collapse can cause massive 
cores to form far away from the centers of molecular clouds on much shorter timescales than the global dynamical collapse time.
Taking into account the existence of YSO clusters and massive clumps at both the EFS ends, 
the observed results are in agreement with the outcome of a model of the end-dominated collapse caused by the higher accelerations of gas. 
Presently, we don't have optically thin line data (such as, NH$_{3}$ and CS) for the EFS, hence, the high-resolution molecular line data will be 
required to further explore the EFS and its velocity structure.

It has been observationally evident that numerous complex processes involved in star formation operate in a given star-forming complex \citep[e.g.,][]{kang10,dewangan16}.
At least three YSO clusters are found at one EFS end and 
also appear to be found near the edges of the shell-like morphology linked with the S242 H\,{\sc ii} region (see Figures~\ref{uf6}a,~\ref{uf7}b, and~\ref{uf8}a). 
Hence, the detection of YSO clusters surrounding the S242 H\,{\sc ii} region also indicates the applicability of 
triggered star formation scenario (via an expanding H\,{\sc ii} region) in the S242 site. 
The dynamical or expansion age of the H\,{\sc ii} region is computed to be $\sim$0.5 Myr (for n$_{0}$ = 10$^{3}$ cm$^{-3}$) (see Section~\ref{subsec:hii}). 
The average lifetimes of Class~I and Class~II YSOs are estimated to be $\sim$0.44 Myr and $\sim$1--3 Myr, respectively \citep{evans09}. 
Considering these typical ages of YSOs and the dynamical age of the S242 H\,{\sc ii} region, it appears 
that the S242 H\,{\sc ii} region is too young to trigger further star formation. Hence, the star formation in the S242 site is unlikely 
influenced by the S242 H\,{\sc ii} region. 

\section{Summary and Conclusions}
\label{sec:conc}
In the present work, we have studied the physical environment, molecular gas distribution, and stellar population in and around the S242 site, 
using the multi-wavelength data. We have chosen a field of $\sim$1$\degr$.05 $\times$ 0$\degr$.56 containing the sources, 
S242, IRAS 05490+2658, IRAS 05488+2657, and IRAS 05483+2728.
The aim of this study is to investigate the physical environment and star formation processes in and around 
the selected target. The major results of our multi-wavelength analysis are the following:\\
$\bullet$ The molecular cloud associated with the S242 site (and three IRAS sources, IRAS 05490+2658, IRAS 05488+2657, and IRAS 05483+2728) is depicted 
in a velocity range from $-$3.25 to 4.55 km s$^{-1}$ and has spatially elongated appearance. \\
$\bullet$ The distribution of ionized emission toward the S242 site detected in the NVSS 1.4 GHz and GMRT 1.28 GHz 
continuum maps and H$\alpha$ image is almost spherical. 
The ionizing photon flux value computed at 1.4 GHz corresponds to a single ionizing star of B0.5V--B0V spectral type. 
The dynamical age of the S242 H\,{\sc ii} region is computed to be $\sim$0.5 Myr (for n$_{0}$ = 10$^{3}$ cm$^{-3}$). 
In the {\it Herschel} column density map, the S242 site is located in the highest column density 
region (peak $N(\mathrm H_2)$ $\sim$2.7~$\times$~10$^{22}$ cm$^{-2}$; A$_{V}$ $\sim$29 mag). \\
$\bullet$ An elongated filamentary structure (EFS) is observed in the {\it Herschel} column density map 
and is embedded within the S242 molecular cloud. 
The S242 H\,{\sc ii} region is located at one of the EFS ends. The highest temperature ($\sim$32 K) is found toward the S242 H\,{\sc ii} region. 
In the temperature map, the EFS is depicted in a temperature range of about 10--12 K (except near the S242 H\,{\sc ii} region). The temperature map traces temperature variations toward the EFS (see areas near both the ends of the EFS).\\
$\bullet$ The EFS has an observed mass per unit length of $\sim$200 M$_{\odot}$ pc$^{-1}$ larger than the 
critical value of $\sim$16 M$_{\odot}$ pc$^{-1}$. 
Eleven {\it Herschel} clumps are found toward the EFS and their masses vary between 150 M$_{\odot}$ 
and 1020 M$_{\odot}$. This implies that the fragmentation has been occurred along the EFS's length. 
The most massive clumps are observed at both the EFS ends and the S242 H\,{\sc ii} region is located at one of the EFS ends.\\ 
$\bullet$ 293 YSOs are identified in our selected field and a majority of these are spatially traced toward the EFS. 
The clusters of YSOs are exclusively found at both the EFS ends, revealing the star formation activities.\\

Taking into account the observational outcomes presented in this paper, the results favour a star formation model of the end-dominated collapse originated by the higher accelerations of gas. 
\acknowledgments 
We thank the anonymous reviewer for constructive comments and suggestions. 
The research work at Physical Research Laboratory is funded by the Department of Space, Government of India. 
This work is based on data obtained as part of the UKIRT Infrared Deep Sky Survey. This publication 
made use of data products from the Two Micron All Sky Survey (a joint project of the University of Massachusetts and 
the Infrared Processing and Analysis Center / California Institute of Technology, funded by NASA and NSF), archival 
data obtained with the {\it Spitzer} Space Telescope (operated by the Jet Propulsion Laboratory, California Institute 
of Technology under a contract with NASA). 
This paper makes use of data obtained as part of the INT Photometric H$\alpha$ Survey of the Northern Galactic 
Plane (IPHAS, www.iphas.org) carried out at the Isaac Newton Telescope (INT). The INT is operated on the 
island of La Palma by the Isaac Newton Group in the Spanish Observatorio del Roque de los Muchachos of 
the Instituto de Astrofisica de Canarias. The IPHAS data are processed by the Cambridge Astronomical Survey 
Unit, at the Institute of Astronomy in Cambridge. 
RD acknowledges CONACyT(M\'{e}xico) for the PhD grant 370405. 
AL acknowledges the CONACyT(M\'{e}xico) grant CB-2012-01-1828-41. 
\clearpage
%

%
\begin{deluxetable}{cccccccc}
\tablewidth{0pt} 
\tabletypesize{\scriptsize} 
\tablecaption{Physical parameters of the {\it Herschel} clumps identified in the region around S242 
(see Figures~\ref{uf4}a and~\ref{uf4}c). Column~1 gives the IDs assigned to the clump. Table also lists 
positions, deconvolved effective radius (R$_{c}$), clump mass (M$_{clump}$), peak column density ($N(\mathrm H_2)$), 
peak temperature (T$_{d}$), and mean central number density (n$_{c}$ = peak $N(\mathrm H_2)$/(2 R$_{c}$)). 
The column density value can also be used to obtain the extinction using the relation $A_V=1.07 \times 10^{-21}~N(\mathrm H_2)$.
The clump IDs 1--11 are distributed toward the EFS containing the S242 site (see Figure~\ref{uf4}c). \label{tab1}} 
\tablehead{ \colhead{ID} & \colhead{{\it l}} & \colhead{{\it b}} & \colhead{R$_{c}$}& \colhead{M$_{clump}$} & \colhead{peak $N(\mathrm H_2)$} & \colhead{peak T$_{d}$} & n$_{c}$\\
\colhead{} &  \colhead{[degree]} & \colhead{[degree]} & \colhead{(pc)} &\colhead{($M_\odot$)} & \colhead{$\times$10$^{21}$ (cm$^{-2}$)}& \colhead{(K)} & \colhead{(cm$^{-3}$)}}
\startdata 
          1\tablenotemark{a}  &    182.453    &        0.197	      &    1.2     &	  260   &6.9&16 &930\\	  
          2\tablenotemark{a}  &    182.403    &        0.247	      &    1.7	   &	 1020   &27 &26 &2575\\
          3  &    182.341    &        0.247	      &    1.4	   &	  420  & 13 &18&1505\\
          4  &    182.177    &        0.239	      &    1.6	   &	  480  &8.3 &14&840 \\
          5  &    182.068    &        0.263	      &    1.5	   &	  380  & 4.7 &14&505 \\
          6  &    181.987    &        0.290	      &    1.6	   &	  405    &5.6 &13&565\\
          7  &    181.971    &        0.302	      &    0.8	   &	  150  & 9 &13 &1820\\
          8  &    181.928    &        0.325	      &    1.2	   &	  350   & 7.2 &13&970\\
          9\tablenotemark{a}  &    181.917    &        0.364	      &    1.5	   &	  700    &15.8 &16&1705 \\
         10\tablenotemark{a}  &    181.851    &        0.309	      &    1.5	   &	  700  & 12.7 &16&1370 \\
         11  &    181.777    &        0.329	      &    1.0	   &	  150   &  3.6 &13&580\\
         12  &    182.030    &        0.364	      &    0.9	   &	  100   &2.8  &12&505 \\
         13  &    182.006    &        0.368	      &    0.8	   &	  125   &5.1 &16&1030 \\
         14  &    182.057    &        0.399	      &    1.1	   &	  250   &6.8&10 &1000\\
         15  &    182.037    &        0.414	      &    1.1	   &	  265   &6.7  &10&985\\
         16  &    181.870    &        0.453	      &    0.5	   &	   25   &1.9 &18&615 \\
         17  &    182.127    &        0.119	      &    0.5	   &	   35   & 2.1 &17&680\\
         18  &    182.193    &       -0.021	      &    0.5	   &	   45  & 3.3 &15 &1070\\
\enddata  
\tablenotetext{a}{It is found at the EFS end.}
\end{deluxetable}

\end{document}